# Programmable skyrmions for robust communication and intelligent sensing

Long Chen [1, 2, †], Xin Yu Li [1, 2, †], Yijie Shen [3, 4, *], Ze Gu [1, 2], Jian Lin Su [1, 2], Qiang Xiao [1, 2], Si Qi Huang [1, 2], Shi Long Qin [1, 2], Qian Ma [1, 2, *], Jian Wei You [1, 2, *], and Tie Jun Cui [1, 2, 5*]

[1]*State Key Laboratory of Millimeter Wave, Southeast University, Nanjing, China.*
[2]*Institute of Electromagnetic Space, Southeast University, Nanjing, China.*
[3]*Centre for Disruptive Photonic Technologies, School of Physical and Mathematical Sciences, Nanyang Technological University, Singapore 637371, Singapore*
[4]*School of Electrical and Electronic Engineering, Nanyang Technological University, Singapore 639798, Singapore*
[5] *Suzhou Laboratory, Suzhou, Jiangsu 215000, China.*

[†]*These authors contributed equally to this work*
[*]*Email: yijie.shen@ntu.edu.sg, maqian@seu.edu.cn, jvyou@seu.edu.cn , and tjcui@seu.edu.cn*

The recently observed plasmonic skyrmions, as electromagnetic counterparts of topologically stable quasiparticles, hold significant promise as novel carriers for robust information transfer and manipulation of nontrivial light-matter interactions. However, their practical applications has been hindered by the lack of flexible tuning devices to encode these topological structures. Here, we present a programmable plasmonic skyrmion platform capable of encoding diverse skyrmion topologies, including Néel-type skyrmions and merons. Based on unprecedented ultra-fast coding feature, we synthesize harmonic skyrmions in the temporal dimension and, for the first time, applied skyrmions in communication and sensing applications. Specifically, we achieved highly robust and multi-channel wireless communications by using programmable topological skyrmions, providing a promising platform for communication in turbulent noise channels and extreme conditions. Furthermore, we implemented intelligent sensing across twenty animal models on the same platform, achieving high recognition accuracy. This methodology offers programmable and temporal insights into the skyrmions for their practical applications in next-generation wireless communication and intelligent sensing.

## Introduction

Skyrmions, as non-trivial topological textures with stable three-dimensional vector field configurations, have attracted considerable attention due to their diminutive size and robust topological stability[1-6], rendering them ideal candidates for compact and resilient information carriers[7-10]. Since Skyrme's initial proposal of their role in describing meson interactions[11,12], skyrmions have been extensively explored across various physical systems, including elementary particles[12], Bose-Einstein condensates[13], nematic liquid crystals[14], magnetic materials[15,16], and twistronics[17]. The discovery of magnetic skyrmions in chiral magnets, observable in real space[18], has notably opened new avenues for high-density data storage and low-energy magnetic storage devices[8,19]. In addition to magnetic skyrmions, optical skyrmions holds significant importance as well. They not only extend the research frontiers of skyrmions into cutting-edge fields of optical domains but also leverages the unique advantages of optical platforms to explore novel physics and applications of skyrmions.

In recent years, electromagnetic (EM) skyrmions have emerged as a pioneering research frontier. Unlike their magnetic counterparts, EM skyrmions offer the advantage of operating at optical and microwave frequencies, thereby extending the skyrmionic phenomena to a broader spectrum of applications. In the optical domain, skyrmionic structures form lattices in specialized surface plasmon polariton (SPP) resonators or appear as stable skyrmions from spin-orbit-coupled SPP fields. This development has significantly expanded the family of skyrmions, allowing various physical quantities to describe them, such as skyrmion textures formed by Poynting vectors[20], tunable skyrmions based on Stokes vectors[21], spin vectors[22], momentum vectors[23] and EM field vectors[4]. These skyrmions present substantial potential for applications across different fields. In the microwave range, electromagnetic skyrmions can be generated via the magnetic and electric fields of spoof localized surface plasmons (SLSP)[24]. These skyrmions offer low energy consumption and deep penetration, making them suitable for high-performance communication and sensing.

Despite their promising features—ultra-small size, ultra-fast speed, topological diversity, and stability—skyrmions have faced significant challenges in achieving effective tunability. The concept of a programmable skyrmion remains an elusive goal not only in electromagnetic systems but also in magnetic systems. This limitation hampers their programmability and, consequently, their practical application potential. Programmability[25-27] is the cornerstone of advanced information storage and processing technologies, enabling flexible switching and dynamic optimization for diverse tasks. However, the current technological landscape lacks mechanisms for stable control of skyrmions under varying external conditions, which is essential for their integration into functional devices. The core challenge is the difficulty in achieving programmable control, as the lack of effective methods to dynamically and reliably manipulate these topological textures has significantly hindered their practical implementation, despite their well-documented theoretical advantages. To better unlock the potential of skyrmions, advancements in programmable control are essential.

In this work, we present a programmable SLSP skyrmion platform. By integrating active elements, we achieve intelligent control over various topological textures, such as Néel-type skyrmions and merons. When all active elements switch off, the out-of-plane and in-plane components of the electric field synthesize into a scalar vortex with topological charge of 0 and a polarization vortex with topological charge of 1, respectively, resulting in a Néel-type skyrmion with a skyrmion number of 1. Conversely, when all elements switch on, a meron is generated, demonstrating unprecedented skyrmion programmability. Notably, by exploiting the ultrafast tuning capability of the active components, we synthesize harmonic skyrmions in the time dimension, representing a stark departure from traditional methods of inducing nonlinearity through material properties.

Furthermore, we apply the skyrmions to communication and sensing fields for the first time. Employing programmability-based binary amplitude shift keying (BASK) and harmonics-based orthogonal frequency-division multiplexing (OFDM) techniques, we achieve robust, low-power communication in ultracompact devices. The topological protection characteristics provide a promising platform for communication in turbulent noise channels and extreme conditions. Additionally, our platform achieves high recognition accuracy of diverse feature information from extensive measured data of twenty animal species, demonstrating functional integration on a single platform. This work paves new avenues for the research and practical application of skyrmion programmability, highlighting its significant impacts in both topological physics and information technology.

## Programmable skyrmion platform

The advent of reconfigurable technology has opened up new possibilities for the dynamic control of EM properties across various frequencies. These advancements have facilitated the integration of mechanical, electrical, thermal, and optical control methods[24-31], leading to the development of highly adaptable platforms. Among these, electrical reconfigurability, which leverages active elements such as positive-intrinsic-negative (PIN) diodes and varactor diodes, offers ultra-fast tuning capabilities that significantly surpass other approaches. Here, we present a programmable SLSP skyrmion platform that leverages these advantages in electrical reconfigurability. Our platform employs PIN diodes within each arm of the SLSP skyrmion structure, enabling precise and independent control over the on/off states of these diodes. This setup allows for a binary state configuration (on: 1, off: 0), yielding $2^8 = 256$ distinct encoding schemes, as depicted in Fig. 1**a**. By analyzing the electric-field textures produced under different encoding schemes, we identified two classical topological textures (i.e., Néel-type skyrmion and meron) arising from two unique encoding configurations. When the PIN diodes are all set to the “off” state (“00000000”), the SLSP skyrmion exhibits perfect $D_8$ point group symmetry, effectively exciting an electric-field-based Néel-type skyrmion[32-34] (coded as 0). Conversely, setting all diodes to the “on” state ("11111111") generates a unique topological texture known as a meron[35-37] (coded as 1). The structural parameters of the SLSP skyrmion, including the length and curvature of the sector arms, the positioning of the PIN diodes, and the substrate radius, predominantly determine its spectral and field confinement characteristics. To optimize these parameters, we conducted extensive numerical simulations. The programmability of our platform extends beyond static configurations. By employing a specific time-periodic driving protocol, we induce novel harmonic phenomena. For instance, Fig. 1**b** illustrates that this protocol can generate Néel-type skyrmions of varying intensities at harmonic frequencies. Each cycle $T$ of the switching sequence consists of two time-steps, producing synthetic harmonic skyrmions in the time dimension. Given the open system nature and dispersion of harmonic modes, traditional analytical frameworks fall short. Therefore, we developed a time-varying model to investigate these time-varying harmonic characteristics (see Supplementary Information Note 5 for details).

Furthermore, we applied these unique harmonic characteristics to implement a multi-user frequency division multiplexed communication system, as illustrated in Fig. 1**c**. Users operating at the fundamental frequency $f_0$ and its harmonics $f_n = f_0 \pm nf_s$ $(n = 1,2,\cdots)$ can achieve high-quality communications, including images and videos, while users at other frequencies cannot exchange information. This topologically protected information transmission ensures robustness against obstacles and interference. In addition to communication[38-40], we explored the robust transmission properties of Néel-type skyrmions for sensing applications. By integrating artificial intelligence (AI), in particular convolutional neural networks (CNNs), we achieved high accuracy in classifying and identifying twenty different animal models based on the amplitude, phase, and complex information (amplitude-phase) of the electric field vector. Figure 1**d** shows that our platform can achieve recognition accuracies exceeding 98%. Comparative analyses with backpropagation and recurrent neural networks further underscore this high accuracy, as detailed in the Supplementary Information. In conclusion, our programmable SLSP skyrmion platform not only demonstrates high programmability and flexible control over diverse topological textures, but also reveals important applications in robust communication and intelligent sensing.

## Spatial and temporal topological multiplexing

In this section, we demonstrate the dynamic control of topological properties across both spatial and temporal dimensions using our innovative programmable SLSP skyrmion platform. This dual-dimension control is instrumental in advancing the utility of skyrmions in communication and sensing. Dynamic modulation in the spatial dimension is achieved through eight PIN diodes integrated within the SLSP structure[41], as illustrated in Fig. 2**a**. Note that this spatial modulation does not refer to the traditional concept of spatial modulation at distinct positions; rather, it involves modulation transformations of the electric field configuration within the same region. The blue and red colors indicate the off and on states of the diodes, respectively. The first layer of the SLSP resonator comprises a copper film etched onto a dielectric substrate, with the third and fifth layers made from the same dielectric material (see Fig. 2**b**). This multi-layer configuration ensures optimal resonance and EM field confinement, which is critical for effective skyrmion generation. Our single-port feed network design represents a streamlined and efficient method to excite the desired skyrmions, providing an intuitive understanding of the symmetry matching condition. Detailed fabrication methods and additional experimental configurations are provided in the Supplementary Information and Methods section. The binary states of the PIN diodes allow for 256 distinct coding strategies (see Fig. 2**d**), offering extensive flexibility and rapid programmability in the spatial dimension. By strategically switching the diodes, we can generate various topological textures. For instance, when all diodes are switched off, an SLSP forms at the resonant frequency, as shown by the reflection coefficient $S_{11}$ in Fig. 2**e**. At a height of $z = 10$ mm above the structure, the phase distribution of the electric field components $E_R$ and $E_L$ indicates topological charges of -1 and +1,[42] respectively, as shown in Fig. 2**e**, with the corresponding magnitude and phase of the electric field components along *x*, *y*, and *z* directions also depicted. The configuration of the electric field vectors shows an "upward" direction at the center, gradually flipping to "downward" at the edges (refer to Fig. 2**g**). By calculating the skyrmion number $S$ (as detailed in Supplementary Note 4, where we provide the integration regions for the skyrmion number calculations for two different textures), we experimentally confirmed this field configuration as a Néel-type skyrmion. Conversely, when all PIN diodes switch on, a different topological quasiparticle—a meron—is generated. The magnitude and phase of the electric field components are shown in Fig. 2**f**, with the corresponding texture depicted in Fig. 2**h**. It is important to note that while different coding schemes produce various electric field textures, this work focuses specifically on these two topological textures for robust communication using 1-bit information encoding.

Beyond spatial modulation, our platform enables dynamic control in the temporal dimension, facilitating the synthesis of harmonic skyrmions. Time-varying systems present new opportunities to elucidate harmonic phenomena, but the simulation of time-varying systems remains a formidable challenge. To overcome this, we employed a time-varying modeling strategy and achieved accurate simulations of harmonic skyrmions (see Supplementary Information for details). We investigated the generation of these harmonic skyrmions by applying a monochromatic signal with a pump frequency of $f_0 = 4.358$ GHz and adopting a square wave driving protocol. Figure 2$\mathbf{j}$ shows the response at a switching frequency of $f_s = 100$ MHz. The output signal intensity distribution reveals additional peaks at different harmonics of the switching frequency $f_s$, corresponding to frequencies $f_n = f_0 \pm nf_s$ $(n = 1,2,3,\cdots)$, indicative of harmonic Néel-type skyrmions. Meanwhile, the measured spatial distribution of the $z$-component of the electric field $|E_z|$ for these harmonic skyrmions, presented in Fig. 2**k**, reveals a trivial scalar vortex mode. The generation of these harmonic skyrmions leverages the open-system nature of our programmable platform, where energy is injected into the system through time-varying modulation. This mechanism fundamentally differs from traditional methods that rely on the inherent nonlinearity of materials to generate harmonic waves in static systems. For more detailed information on the numerical simulations and experimental tests (refer to the Supplementary Information). The combined spatial and temporal control of skyrmions via programmability introduces a novel mechanism for exploring exotic physical phenomena that not only enhances our understanding of topological matter, but also holds significant potential for practical applications in advanced communication and sensing.

## Skyrmion-based robust wireless communication

Topological photonic crystals safeguard their unique features through topological invariants defined in reciprocal space, whereas skyrmions manifest their topological properties in real space. This distinction paves the way for extensive applications of topological states in information processing and transmission. Leveraging the programmable attributes of the proposed SLSP skyrmions, we developed a wireless communication system based on BASK modulation. The conceptual diagram of the experimental setup is illustrated in Fig. 3**a**, which enables straightforward information transmission. The photograph of the experimental setup for the BASK communication system is depicted in Fig. 3**b**. Using a microcontroller, the sender encodes an image into a binary bit stream by switching the PIN diodes to regulate the skyrmions between the states representing code 0 and code 1. A receiving antenna, located 20 cm apart from the transmitter, captures the signal intensity, which is then processed for image reconstruction. Figure 3**c,d** presents the transmission results for the black-and-white logo of *Southeast University* and the color logo of *Nanyang Technological University*, respectively (see Methods and Supplementary Information for details on signal detection). The constellation diagram of the BASK communication is shown as a histogram, confirming the high-quality performance of our system.

To further improve communication speed, we developed an advanced wireless communication system by maintaining robust Néel-type skyrmion texture without switching the PIN diodes. The schematic of the system is shown in Fig. 3**e**. At the receiving end, a software-defined radio transceiver receives signals from the skyrmions. The quality metrics of the received video, illustrated in Fig. 3**f**, demonstrate the robustness of the OFDM communication system based on SLSP skyrmions, with the peak signal-to-noise ratio (PSNR) consistently ranging from 40 to 60 dB, and the structural similarity index metric (SSIM) exceeding 99%. Furthermore, the accuracy of the video transmission is highlighted in Fig. 3**g**, underscoring the effectiveness of this approach (see Supplementary Video 3 for details), while specific experimental setups can be found in the Methods section. To demonstrate the multi-frequency processing capability, we exploited the harmonic characteristics of the system. Experimental results indicate that users at different harmonics can achieve high-quality image reconstruction (see Fig. 3**h**), enabling frequency division multiplexing communication. However, the current hardware platform limits us to transmit identical information (see Supplementary Note 12). One of the most attractive advantages of our communication system is its topological protection. The experimental results show that two traditional antennas are unable to achieve high quality line-of-sight communication in the presence of obstacles, resulting in noticeable image distortion. However, replacing the traditional transmitting antenna with our developed robust antenna based on SLSP skyrmions allows for high-resolution image reconstruction (see Methods and Supplementary Note 24 for specific experimental details). This topologically protected information transmission vastly surpasses the anti-interference capabilities of traditional systems. Thus, we promoted the important application of skyrmions in wireless communications. This may provide a promising solution for communications in turbulent noise channels and extreme conditions, such as disaster-stricken areas with collapsed structures or densely packed urban centers with numerous skyscrapers, where robustness is crucial. However, it should be noted that the size of obstacles is not unrestricted and is related to the dimensions of the SLSP skyrmion transmitting antenna, as detailed in Supplementary Note 22. Additionally, the current system, being a surface wave array, is limited to near-field communications over distances of several tens of centimeters. Future advancements may use metamaterial structures to extend the capabilities to far-field communication.

## Skyrmion-based intelligent sensing

While we have demonstrated the impressive capabilities of the programmable SLSP skyrmion platform in communication applications, its potential is vast and extends far beyond the confines of these initial demonstrations. The same robustness that benefits communication systems also lay the foundation for advanced sensing technologies. By combining the unique topological properties of skyrmions with the powerful learning capabilities of AI, we can develop innovative intelligent sensing solutions that outperform traditional wireless sensing methods, which are highly susceptible to environmental interference. By contrast, skyrmions exhibit remarkable resistance to such interference, making them ideal for robust sensing applications. Here, we propose a novel method for intelligent sensing using programmable SLSP skyrmions, as illustrated in Fig. 4**a**. When different animal models are placed in the skyrmion propagation path, the interaction between the skyrmions and the models influences the EM response. A vector network analyzer (VNA) measures and collects these EM-response data, which are subsequently used to train a CNN for accurate animal model identification. The overall structure of the sensing system is presented in Fig. 4**b**. The experimental platform primarily consists of the programmable SLSP skyrmions, a VNA, a receiving antenna, a near-field scanning microwave microscopy (NSMM), and a computer. To enhance data diversity, we made adjustments to the position and posture of each animal model and collected multiple datasets. Figure 4**c** showcases 20 different animal models, labeled sequentially from 0 to 19. The animal models were placed 10 cm in front of the developed robust antenna based on SLSP skyrmions. Measurement data (see Supplementary Information) indicated that these different animal models exhibit distinct response characteristics at various frequency points within the test bandwidth, ensuring excellent classification performance. In this experiment, we collected data at 501 evenly spaced frequency points in the range from 4 to 4.5 GHz.

Based on the collected experimental datasets, we designed a CNN, with detailed architecture provided in the Supplementary Information. The input to the CNN consists of an EM-response data matrix including 36 spatial elements ($6 \times 6$), and the output layer corresponds to 20 categories. It is noteworthy that we selected the central point of the metallic pattern of the skyrmion sample as the reference center in the experiment, and positioned 36 points at spatial intervals of 20 mm. At each frequency point, a spatial distribution matrix could be obtained, and increasing the number of frequency points would yield a corresponding multiple of input data. Our network achieved an accuracy rate of 98.33% during training. The accuracy and loss curves, as well as the confusion matrix for the test dataset, are depicted in Fig. 4**d,e**. Moreover, the probability distribution of a successful prediction across different labels is illustrated in Fig. 4**f**. We further analyzed the impact of various parameters on recognition accuracy. As shown in Fig. 4**g**, increasing the size of the EM-response data matrix enhances performance, with accuracy stabilizing above 98%. Variations in accuracy with different frequency points are presented in Fig. 4**h**, indicating a significant decline in identification accuracy when frequency points deviate from the resonance frequency. This is primarily due to the highest conversion efficiency of skyrmions occurring at the resonance point, where the energy received on the same cross-section is strongest, and the scattering field differences between different objects become more pronounced. Conversely, deviations from the resonance frequency result in reduced conversion efficiency and lower received energy, complicating system recognition. Additional analyses in the Supplementary Information cover the learning accuracy of different neural networks and the impact of different data types on recognition accuracy. Moreover, we analyzed the effects of sensing distance and the size of recognized objects on the system's sensing and recognition accuracy (see Supplementary Note 25 for details). By integrating these functionalities on the same platform, we not only achieve efficient communication but also pioneer a new dimension of intelligent sensing through the unique topological properties of SLSP skyrmions.

## Discussion

In this work, we have demonstrated a programmable plasmonic skyrmion platform that enables dynamic control over diverse topological textures, including Néel-type skyrmions and merons, through spatial and temporal modulation. By integrating active elements such as PIN diodes, we achieved unprecedented programmability, synthesizing harmonic skyrmions in the temporal dimension and pioneering their application in robust wireless communication and intelligent sensing. The platform leverages topological protection to deliver high-fidelity data transmission in turbulent environments and achieves exceptional recognition accuracy in multispecies classification via AI integration. These advancements bridge topological physics with practical applications, offering a versatile framework for future innovations. We remark that the SPP field only supports TM modes, whose electric field lacks azimuthal components during propagation, thus only allows for the generation of radial hedgehog-shaped Néel-type skyrmions.[1,3] To achieve chiral Bloch-type skyrmions, the introduction of specially designed complex anisotropic chiral medium layers and other EM vectors is necessary.[43] Consequently, our current research is solely focused on these two distinctly different topological states: the Néel-type skyrmion and meron. This focus represents a foundational step toward the realization of programmable skyrmion applications. However, the potential for future development remains substantial. The skyrmions exhibit a rich diversity of the topological states, which could extend our platform from the unit level to array level, enabling the realization of more possible topological textures through metamaterials. This would not be limited to Néel-type skyrmions and merons, but could also expand the current 1-bit information encoding communications to high-dimensional coding communications. By incorporating the concept of time, we could advance to use spatiotemporal metamaterial platforms to realize spatiotemporal skyrmions and other novel physical phenomena. This is expected to significantly enhance the scope of physical exploration and practical applications. By addressing these challenges and opportunities, our work lays a foundation for next-generation topological devices that merge physics, engineering, and artificial intelligence.

## Acknowledgements:

This work was partially supported by the National Natural Science Foundation of China (62288101, 92167202), the National Key Research and Development Program of China (2023YFB3813100), Nanyang Technological University Start Up Grant, Singapore Ministry of Education (MOE) AcRF Tier 1 grant (RG157/23), MoE AcRF Tier 1 Thematic grant (RT11/23), Singapore Agency for Science, Technology and Research (A*STAR) MTC Individual Research Grants (M24N7c0080), Special Fund for Key Basic Research in Jiangsu Province (Nos. BK20243015), the Young Elite Scientists Sponsorship Program by CAST (2022QNRC001), the State Key Laboratory of Millimeter Waves, Southeast University, China (K201924), the 111 Project (111-2-05), and the China Postdoctoral Science Foundation (2021M700761, 2022T150112).

## Author contributions:

Y.S., J.Y. and T.C. conceived the idea. L.C., Z.G., X.L. and Q.M. designed the experiments. Y.S., J.Y. and T.C. supervised the project. L.C., Z.G., X.L., J.S., Q.M. and Q.X. conducted the experiments, collected and analyzed the data. L.C., Y.S. and J.Y. carried out the simulations, theoretical analyses, and wrote all the code. L.C., Y.S., J.Y. and T.C. wrote the manuscript, with contributions from all authors.

## Competing interests:

The authors declare no competing interests.

## Data and materials availability:

All data needed to evaluate the conclusions in the paper are present in the paper and the Supplementary Materials.

# METHODS

## Numerical simulations of the time-varying system

We performed numerical simulations of the time-varying system using CST Microwave Studio, a commercial electromagnetic simulation software. The simulations were initiated by driving the excitation port with a continuous wave single-tone signal, lasting between 10 to 20 cycles, at a frequency corresponding to the resonance frequency $f_0 = 4.319$ GHz when all switches were in the OFF state. This resonance frequency was critical for ensuring accurate system behavior under static conditions. To capture the near-field results, an electric field monitor was placed 10 mm above the surface of the programmable SLSP skyrmion metal patch. The monitored area was a square region measuring $240 \times 240\ \mathrm{mm}^2$, centered precisely on the midpoint of the SLSP metal patch. This configuration, exhibiting $D_8$ group symmetry, was crucial for evaluating the electromagnetic properties of the system. The dynamic behavior of the simulation was controlled by adjusting the conductivity of the medium, representing the time-varying states of the PIN diode switches. Conductivities were set to $5.8 \times 10^7\ \mathrm{S/m}$ for the ON state and $0\ \mathrm{S/m}$ for the OFF state. These settings were input through the material properties interface, enabling the simulation of various modulation schemes. When a 0/1 binary square wave modulation was applied, the system exhibited harmonic skyrmion formation at harmonic frequencies $f_n = f_0 \pm nf_s\ (n = 1,2,3,\cdots)$. These frequencies were clearly observable in the electric field monitor, demonstrating the system's ability to generate and manipulate skyrmions dynamically.

## Numerical simulations of various topological textures

In the static scenario, the numerical simulation of the programmable SLSP skyrmion continues to be conducted using the commercial software CST Microwave Studio. In contrast to the time-varying dynamic case, the excitation port here was driven by a broadband pulse signal covering a frequency range of 2-5 GHz. This setup was crucial for evaluating the performance of the system performance across a wide spectrum of frequencies. To analyze the electric field vector distribution under different coding schemes, near-field results were obtained using an electric field monitor positioned 10 mm above the surface of our proposed programmable SLSP skyrmion metal patch, which exhibits $D_8$ group symmetry. The monitored area was a square region with dimensions of $240 \times 240\ \mathrm{mm}^2$, centered precisely at the midpoint of the SLSP metal patch. This precise configuration was essential for capturing the intricate details of the electromagnetic field distributions. For these numerical simulations, we employed real PIN diodes (MADP-000907-14020) and real inductors (LQW15AN22NG80D), incorporating their respective S2P files to ensure accurate representation of their electrical characteristics. This meticulous approach allowed us to simulate the physical behavior of the circuit components with high fidelity, thereby providing more reliable and realistic simulation results.

## Sample fabrications

Our programmable SLSP skyrmion sample was meticulously fabricated using a multilayer lamination process based on Printed Circuit Board (PCB) technology. As illustrated in Fig. 2b, the first layer comprises a microwave plasmonic resonator made from copper film, printed on a Rogers 4530B dielectric substrate. This substrate features a relative permittivity of 3.48, a loss tangent of 0.0037, and a thickness of 0.508 mm, optimizing the resonator's performance. The resonator has an inner radius $r_3$ of 12 mm and an outer radius $r_1$ of 52 mm. The positions of the PIN diodes correspond to an inner radius $r_2$ of 20.85 mm, with a slot width of 0.3 mm, strategically designed to enhance the functionality of the system. The third layer is etched with a microstrip line of length $l = 5.1$ mm, crucial for signal transmission. Both the fourth and sixth layers are composed of the same Rogers 4530B dielectric substrate used in the second layer, ensuring uniformity and stability throughout the structure. The fifth layer, also made from copper film, serves as the grounding plane for the microstrip line. It features a central circular aperture, eight symmetrically arranged circular apertures, and an additional circular aperture for grounding (negative electrode), with radii $r_4$ of 0.75 mm and $r_5$ of 1.5 mm, and positional radii $r_6$ of 23.96 mm and $r_7$ of 15.49 mm. The seventh layer comprises the bias line and feed network layers. The feed microstrip line has a length $l_2$ of 50.59 mm, designed to match an input impedance of 50 Ω, making it compatible with standard SMA connectors. Eight positive bias lines extend to lengths $l_1$ of 29.3 mm, while the negative bias line measures $l_3$ at 39.2 mm. The width for connecting the SMA coaxial connector is precisely 13 mm, ensuring a secure and efficient connection. Commercially available PIN diodes (MADP-000907-14020) and inductors (LQW15AN22NG80D) were utilized during the manufacturing process. These components were chosen for their reliability and performance, ensuring the overall robustness of the fabricated sample.

## Experimental measurements

In our investigations of the programmable and harmonic properties of skyrmions, we employed near-field scanning microwave microscopy (NSMM) to measure near-field electric characteristics and transmission properties. This advanced setup enabled us to capture detailed spatial electric field distributions and transmission characteristics with high precision. The experimental system comprised a vector network analyzer (Agilent N5230C) and several phase-stable coaxial cables. Two coaxial cables were connected to the analyzer's ports: one served as the excitation source linked to the SMA coaxial connector of the programmable SLSP skyrmion sample, and the other was connected to a coaxial probe mounted on a movable platform for detecting spatial electric field distributions. By adjusting the probe's orientation in the x, y, and z directions, we were able to measure all vector components of the electric field $(E_x, E_y, E_z)$. The probe was mounted on a scanning platform, allowing point-by-point measurements to obtain the spatial electric field pattern of a specific area. The probe tip was maintained at a 10 mm distance from the sample surface during testing, with a scanning area consistent with numerical simulations, measuring $240 \times 240$ mm$^2$.

For static programmability testing, standard power supplies were sufficient to switch the coding states. For harmonic testing, the power supply was replaced with a signal generator (Agilent 33600A) to produce the required square wave modulation. This setup allowed for adjustable waveform and duty cycle parameters, with a maximum output voltage of 600 mV. Input excitation was converted from the vector network analyzer (VNA) to a continuous wave single-tone signal generated by a microwave signal source (Keysight E8267D). The receiving probe was connected to a spectrum analyzer (Keysight E4447A) to capture harmonic scattering signals. During this process, the relative position of the probe and the scanning area remained unchanged.

In the BASK-based wireless communication experiment, a microcontroller was employed to control the switching state of eight PIN diodes. The excitation signal was provided by the microwave signal generator (Keysight E8267D). The receiving end utilized a standard broadband Vivaldi antenna, and the received voltage bitstream was obtained via an amplifier and detector. When the coding state is 0, the detected signal corresponds to a Néel-type skyrmion, with a detector voltage of approximately 0.7 V—significantly higher than the 0.2 V detected when the coding state is 1. For the OFDM-based frequency division multiplexing communication system, the microcontroller was replaced with a signal generator (Agilent 33600A) to achieve high-speed switching and harmonic generation. Signal reception was conducted through a universal software radio peripheral device (USRP-2974). In this experiment, the distance between the receiver and the skyrmion transmitter is maintained at 20 cm, consistent with BASK communication. Given that the experiment is conducted entirely within a closed normal indoor environment, the noise level is expected to be less than or equal to -110 dBm/100 kHz, meeting standardized criteria[44]. In the robust communication experiment, the transmitter consists of our skyrmion samples, with the switching states still controlled by the microcontroller. The receiving end comprises an antenna array positioned 20 cm away from the transmitter. A metal obstacle measuring 8 cm in length, 6 cm in height, and 3 cm in thickness is placed at the center between the transmitter and receiver. The receiving antenna array is connected to a switch matrix (Siglent SSM5144A) for rapid data acquisition, with both data transmission and reception occurring on a PC. In the comparative experiment, specifically the traditional transceiver antenna experiment, both the transmitter and receiver utilize broadband Vivaldi antennas, maintaining a distance of 20 cm, with an identical obstacle placed at the center. The transceiver antennas are connected to a USRP (USRP-2974) transceiver for direct communication. In the smart sensing recognition experiment, the object under test was placed 10 cm in front of the skyrmion, with a scanning grid consisting of 36 ($6 \times 6$) points, centered on the metallic pattern of the skyrmion. Spatial sampling is performed uniformly at 20 mm intervals between adjacent points. At each frequency point, a spatial distribution matrix comprising 36 points is present. The frequency sampling range is from 4 to 4.5 GHz, with a total of 501 evenly spaced frequency points collected.

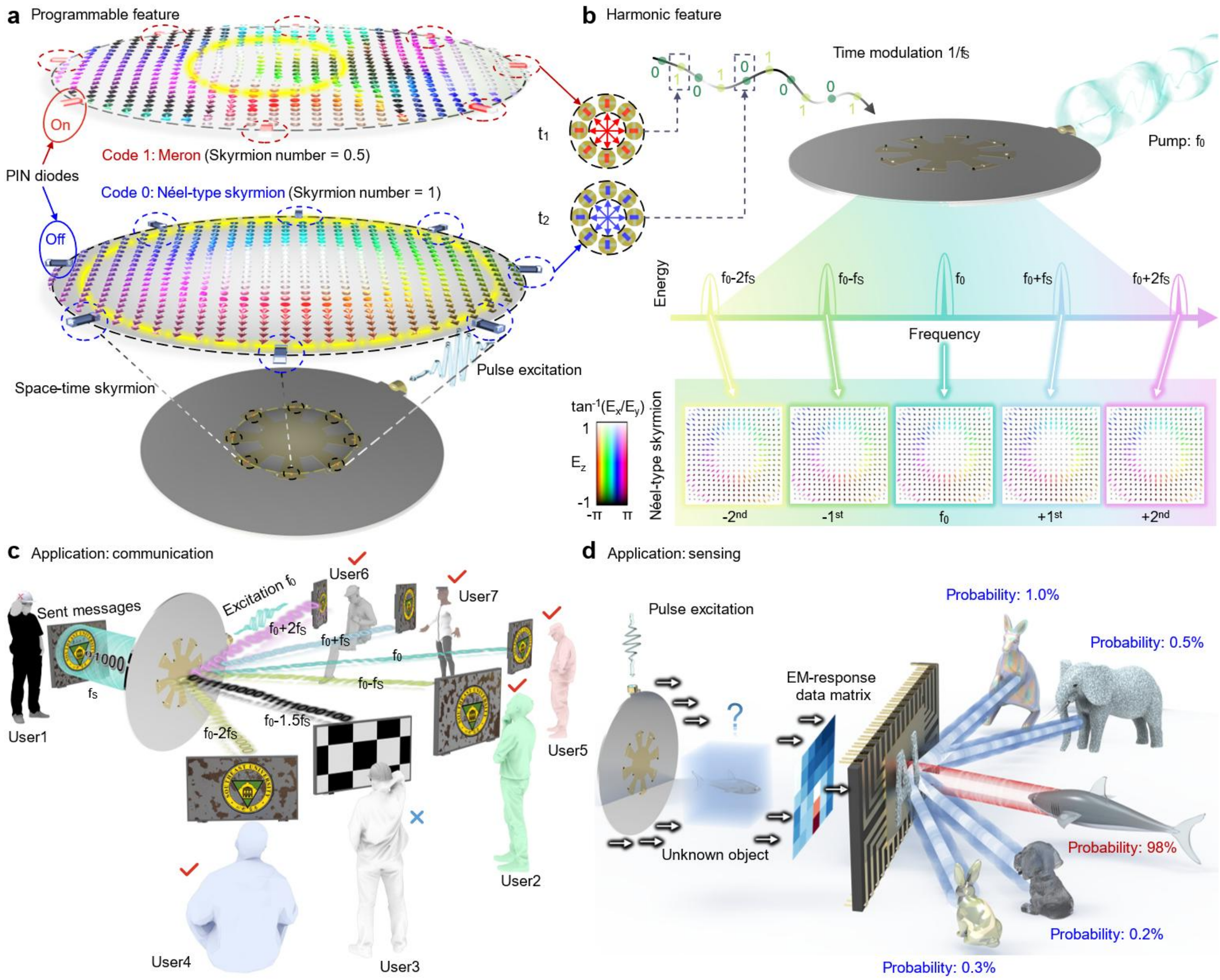

**Fig. 1 | Concept of programmable plasmonic skyrmions for robust communication and intelligent sensing applications. a** The programmable characteristic of two topological textures is controlled by the states of eight PIN diodes. When all PIN diodes switch on (denoted as 1), a meron is supported. Conversely, when all PIN diodes switch off (denoted as 0), a Néel-type skyrmion with a skyrmion number of 1 is generated. **b** Harmonic phenomena induced by the switching frequency ($f_s = 1/T$) of the PIN diodes under a square wave modulation protocol. Harmonic frequencies centered around the original resonance frequency are generated at any switching frequency, and Néel-type skyrmions with a skyrmion number of 1 appear at these frequencies. The mode strengths at different harmonic frequencies can be dynamically controlled by the modulation protocol. **c** A multi-user frequency division communication system leveraging harmonic characteristics. Transmission information, such as images and videos, can be received only at the fundamental and harmonic frequencies generated by harmonic effects, while no information is obtainable at other frequencies. **d** An intelligent sensing and recognition system based on Néel-type skyrmions, combined with artificial intelligence. The recognition accuracy can exceed 98%.

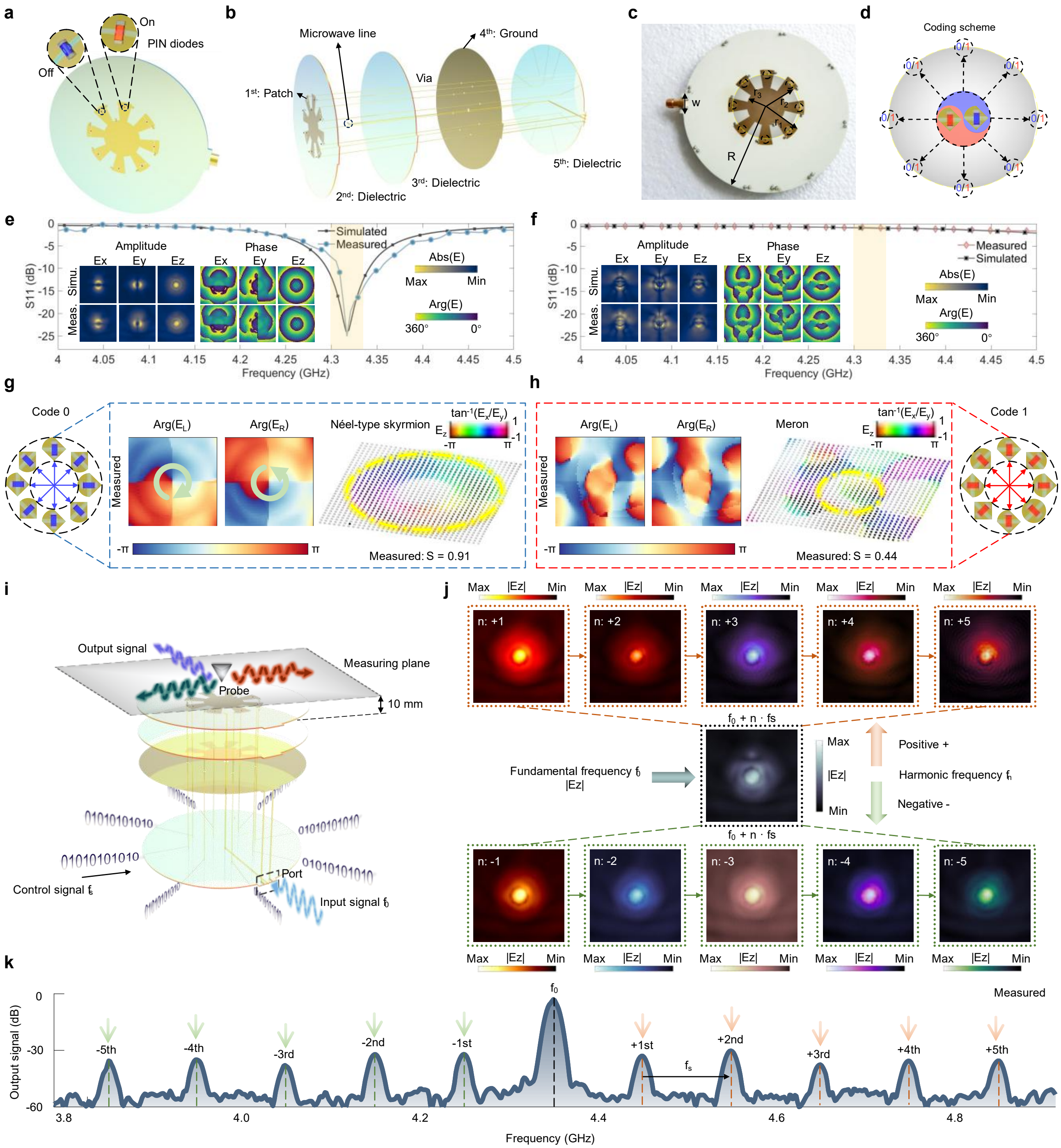

**Fig. 2 | Dynamic properties of skyrmions under spatial and temporal multiplexed modulation.** **a** Schematic of the SLSP skyrmion platform, illustrating the programmable characteristics enabled by PIN diodes. **b** Stratified layers comprising the top patch layer, microstrip line layer, ground layer, feed layer, and three dielectric layers that interconnect these metallic layers. **c** Photograph of the fabricated sample in a frontal view. **d** Description of the encoding scheme: each diode can be in an "on" or "off" state, allowing for 256 potential encoding combinations. **e,f** Numerical simulations and experimental results comparing the reflection coefficient $S_{11}$ and the amplitude and phase distributions of the three electric field components $\{E_x, E_y, E_z\}$ for the configurations where all PIN diodes are off (Code 0: **e**) and on (Code 1: **f**). **g,h** Experimental results of phase and electric field vector distributions for $E_R$ and $E_L$ under two coding states. When all PIN diodes are off (Code 0: **g**), $E_R$ and $E_L$ correspond to scalar

vortex modes with topological charges $l_R = 1$ and $l_L = -1$, forming a Néel-type skyrmion with hedgehog texture and a skyrmion number $S$ close to 1. When all diodes are on (Code 1: **h**), a meron with $S$ approaching 0.5 is achieved. The hue indicates the in-plane spin orientation angle, while brightness variation corresponds to the out-of-plane spin component. **i** Schematic of harmonic skyrmion generation. **j** Experimental results of the electric field response (output signal) in the frequency domain. The output signal exhibits distinct peaks at $f_n = f_0 \pm nf_s$ ($n = 1,2,3,\cdots$), which can be dynamically adjusted via driving protocols. **k** Experimental results of the electric field distribution ($|E_z|$) for harmonic skyrmions, with a pump frequency of $f_0 = 4.358$ GHz. Harmonic skyrmions are generated at frequencies $f_0 \pm f_s$, $f_0 \pm 2f_s$, $f_0 \pm 3f_s$, $f_0 \pm 4f_s$, and $f_0 \pm 5f_s$, with a driving frequency of $f_s = 100$ MHz.

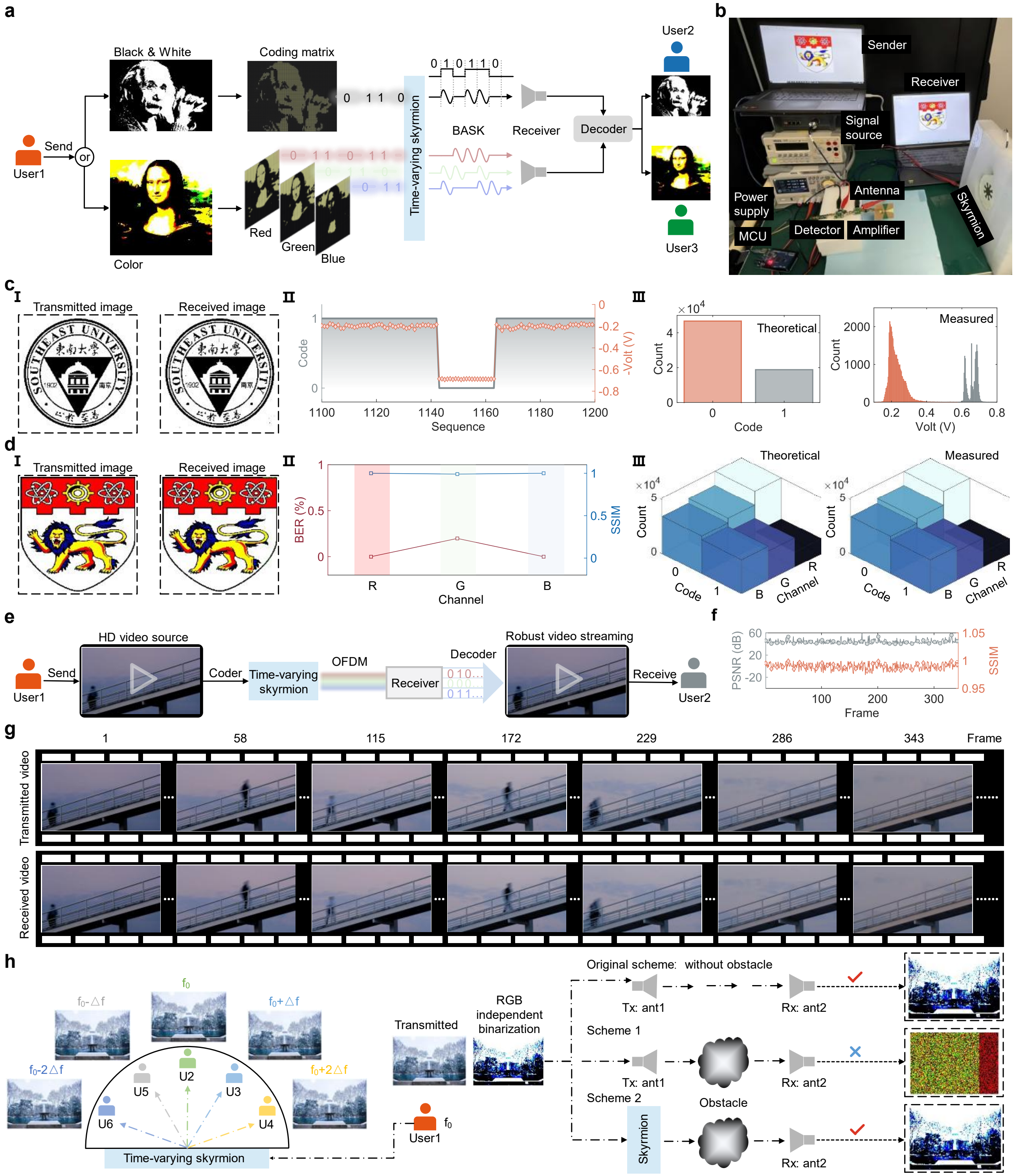

**Fig. 3 | Experimental demonstration of wireless communication systems based on the programmable skyrmion transmitter. a** Conceptual schematic of the experimental setup utilizing BASK modulation. Binary encoding toggles between 0 and 1 to emit radio waves, dynamically transmitting Néel-type skyrmions and merons, which are subsequently captured by users. User 1 transmits black-and-white image data to user 2 via a single channel, while color image data is transmitted to user 3 through three separate channels. Two independent receiving users are positioned at a distance of $z = 50\ cm$ from the skyrmion transmitter. **b** Photograph of the experimental setup.

The wireless communication system comprises a skyrmion transmitter, a receiving antenna, an amplifier, a detector, a microcontroller, a signal generator, and a power supply. **c** Results of black-and-white image transmission using BASK modulation: (I) Transmitted and received images; (II) Collected voltage signals of approximately 100 bits in length and the actual bitstream; (III) Theoretical and measured constellation diagrams for the black-and-white images. **d** Results of color image transmission using BASK modulation: (I) Transmitted and received color images; (II) Quality metrics for the received images in RGB channels—PSNR and SSIM; (III) Theoretical and measured constellation diagrams for the color images. **e** Conceptual schematic of the experimental setup utilizing OFDM modulation. **f** PSNR and SSIM of each frame received by the OFDM-modulated communication system. **g** Comparison of transmitted and received video frames at different frame sequences. **h** Results of a multi-user frequency division multiplexing communication system based on harmonic characteristics, and a comparison of system robustness tests.

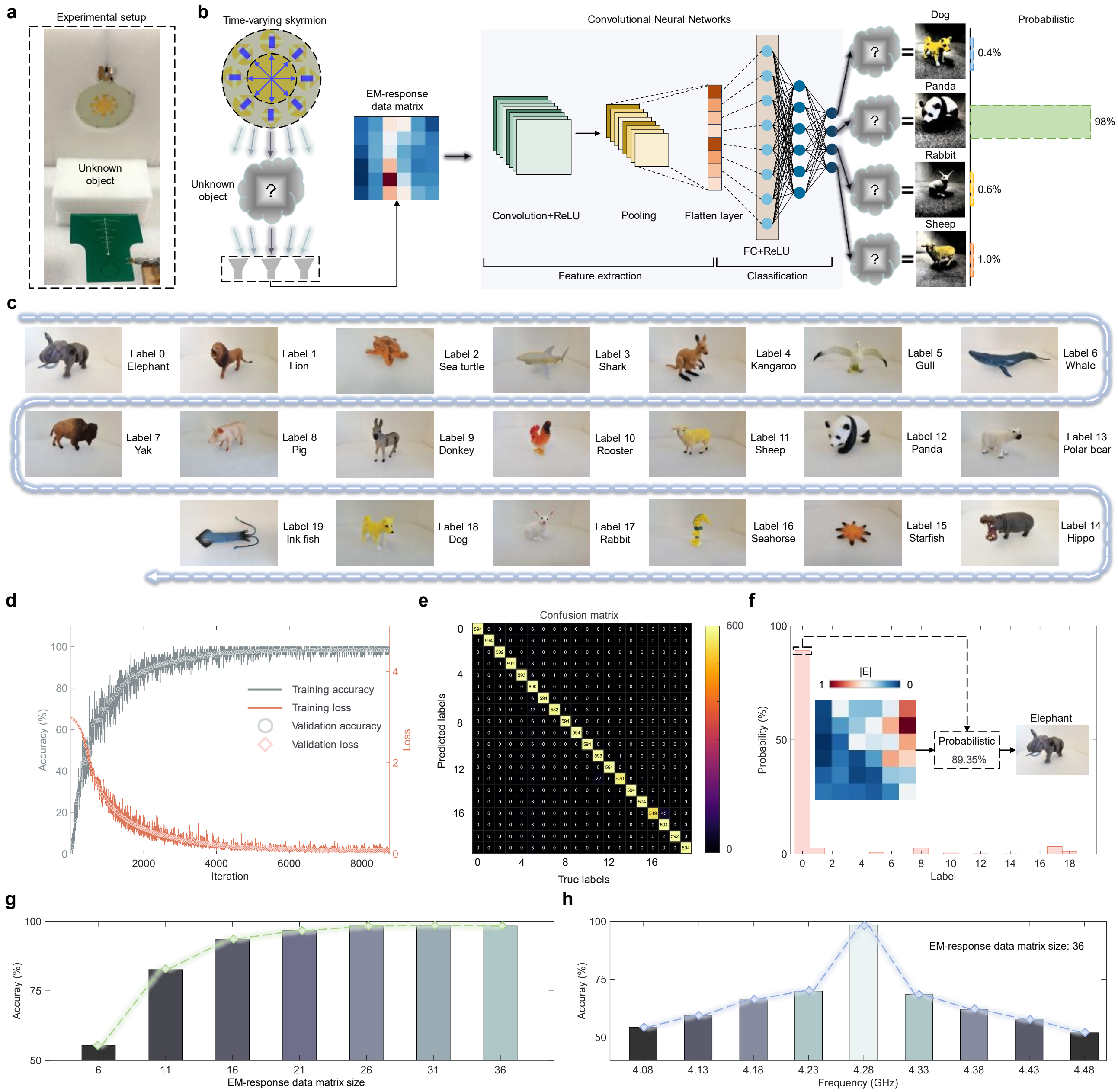

**Fig. 4 | Experimental demonstration of the target intelligent sensing and recognition system based on the programmable skyrmion platform. a** Experimental setup for intelligent sensing and recognition leveraging programmable SLSP skyrmions. The setup comprises a computer, VNA, receiving antenna, NSMM, and the object under test. **b** Schematic illustration of the intelligent sensing and recognition system. The receiving antenna collects scattered signals via the NSMM and processes the energy into the required EM-response data matrix. This EM-response data matrix is then input into a pre-trained CNN, which identifies the target object based on output probabilities. **c** Images of twenty different animal models to be classified, including elephants, lions, and turtles, etc., annotated sequentially from zero to nineteen. **d** Trends of classification accuracy and training loss during the training process. **e** Confusion matrix for the classification task involving 12000 test samples, demonstrating a classification accuracy exceeding 98%. **f** Probability distribution for successfully recognizing a specific elephant test sample across the twenty different labels. **g** Recognition accuracy performance when the EM-response data matrix size changes. **h** Recognition accuracy performance

at different frequencies, maintaining an EM-response data matrix size of 36, with the highest accuracy achieved at the resonant frequency of 4.28 GHz.